\documentclass[a4paper]{article}
\usepackage{interspeech2011,amssymb,amsmath,epsfig}
\usepackage{url}
\usepackage{empheq}
\ninept
\def\reg{{\rm\ooalign{\hfil
     \raise.07ex\hbox{\scriptsize R}\hfil\crcr\mathhexbof20D}}}

\title{Keyphrase Cloud Generation of Broadcast News}

\makeatletter
\def\name#1{\gdef\@name{#1\\}}
\makeatother
\name{{\em Lu\'{i}s Marujo$^{1,2,3}$, M\'{a}rcio Viveiros$^4$, Jo\~{a}o P. Neto$^{2,3,4}$}}

\address{$^1$ Language Technologies Institute  / Carnegie Mellon University, Pittsburgh, USA  \\
  $^2$ Spoken Language Laboratory / INESC-ID Lisboa, Portugal\\
  $^3$ Instituto Superior T\'{e}cnico, Lisboa, Portugal\\
  $^4$ VoiceInteraction, Lisboa, Portugal\\
{\small \tt Luis.Marujo@inesc-id.pt, Marcio.Viveiros@voiceinteraction.pt, Joao.Neto@inesc-id.pt}}

\begin{document}
\maketitle
\begin{abstract}
This paper describes an enhanced automatic keyphrase extraction method applied to Broadcast News. 
The keyphrase extraction process is used to create a concept level for each news. On top of words resulting from a 
speech recognition system output and news indexation and it contributes to the generation of a tag/keyphrase cloud of the top news included in a Multimedia Monitoring Solution system for TV and Radio news/programs, running daily, and monitoring 12 TV channels and 4 Radios.
\end{abstract}


\section{Introduction}
There are large amounts of daily produced video and audio information, from TV and Radio channels, that are not searchable and indexed. Nowadays, despite all the available amount of information, users spend an inordinate amount of time in zapping in a constant search for relevant information. The traditional channels with a fixed and generic program alignment are not useful any more. Video-on-demand (VOD) for series, films, and music are becoming the standard access method. The IPTV and the generalization of set-top-box concept are changing the TV experience. Also the concept of search potentiated by Google will be a future request also for TVs users, since is starting to be the standard method to access information. However the user interface is a problem in terms of access. The way to provide users with a good TV experience will be based on recommendation systems.
However to build recommendation systems for TV and Radio we need additional work. The availability of a video on the web is problematic because there is no way to access it through content. A search engine is blind to this kind of data. 

The advances of Automatic Speech Recognition Tools (ASR) enabled the automatic transcription of Broadcast News (BN)~\cite{Meinedo2010}. Standard speech recognition systems generate raw single-case words, without punctuation marks, with numbers written as text, and with many different types of disfluencies.
The generation of the missing information makes this representation format easier to read and understand, and mitigates problems to further automatic processing~\cite{Jones2003}. Capitalization, also known as truecasing, improves human readability, parsing, and NER (Named Entity Recognition). Punctuation marks are useful for parsing, information extraction, machine translation, extractive summarization, and NER. 

The correct identification of the main concepts is one of the bases for automatic Indexation and Segmentation of BN process~\cite{Amaral2008}. 
There is a growing demand for rich interfaces. Tag cloud, sometimes called word cloud, has become a hallmark of Web 2.0 design. 
Tag clouds are weighted renditions of collections of words (tags) that can be used to represent the concepts, in a visually
appealing way to summarize vast amounts of information\cite{Kuo2007}. 
ASR clouds are known to be viable ways of individually representing podcasts~\cite{Tsagias2008}.
The cloud can reflect a high level description of the news. We hope to create a high level description of news using keyphrase extraction. A keyphrase is a set of relevant words or phrases that appear verbatim in a document and that give a brief summary of its content.

Several keyphrase extraction methods have been proposed; these methods can be categorized into simple statistics, linguistic, machine learning, and hybrid.   
N-Grams~\cite{Cohen1995}, word frequency~\cite{Luhn:1957}, TF*IDF~\cite{Salton:1974}, PAT-tree~\cite{Chien:1997} 
fall into the simple statistics.
The linguistic approaches include both lexical and syntactic analysis~\cite{Ercan:2007}. 
The machine learning techniques are usually supervised -- the algorithms learn a model from training data containing manually identified keyphrases and use this model to classify -- SVM~\cite{Zhang2006}, Naïve Bayes~\cite{Medelyan2006}, CRF~\cite{Zhang2008}, C4.5~\cite{Medelyan2010} are known examples. Hybrids~\cite{Medelyan2010} are a combination of two or more approaches.

This paper describes the development of keyphrase extraction method used to create the middle layer of an hierarchical 3 layers representation of news and it is used to generate tag clouds from the top news of the last 6 hours of Portuguese BNs included in a Media Monitoring Solution (MMS) system. It is also explored several features and classifiers for the keyphrase extraction.

This paper is organized as follows: 
Section 2 presents the System Architecture; the description of the new modules included in the MMS system is the nucleus of the Section 3;  the results are describe in Section 4, and Section 5 concludes and suggests future work.

\section{System Architecture}
\begin{figure*}
\centerline{\mbox{\includegraphics[width=19cm]{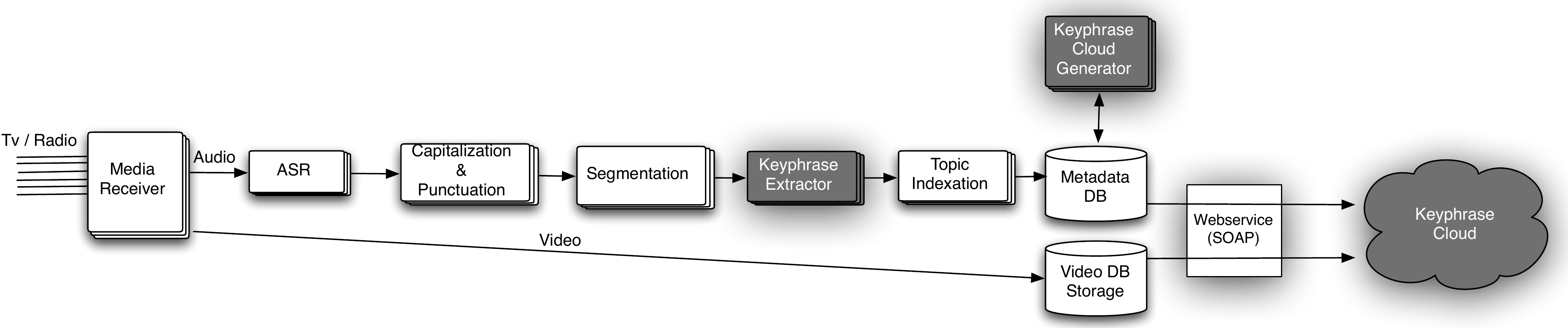}}}
\caption{Component view of the system architecture.}
\label{fig:Architecture}
\end{figure*}
\vspace{-0.2cm}
The figure \ref{fig:Architecture} shows the component view of the MMS system architecture presented in~\cite{Neto2011} where the gray blocks are introduced by this work. The blocks represent software components and the arrows represent the data flow.
The sliced blocks represent several instances of the modules running in parallel. Furthermore, XML is the default internal data representation used between components.

The Media Receiver is responsible for capturing and recording the broadcast news program on TV and Radio in a 24/7 setup. The next component is an ASR that generates the audio transcription based on electronic program guide (EPG) information. Then, the transcription is enriched with Punctuation and Capitalization. Subsequently, each BN program is segmented into several news and  the Keyphrases are extracted -- the extraction process is described in Section \ref{subsection:keyphraseExtraction}. Each news is topic indexed or topic classified. Finally each news is stored in the Metadata DB being constituted by the word transcription, keyphrases, and index, besides program/channel and timing information. 
The Keyphrase Cloud Generator interacts with the Metadata DB creating/updating hourly the keyphrase cloud and storing again the cloud representation in the database. The 3D keyphrase cloud linked to the videos are shown when a user access the system and interact with the databases using a webservice that abstracts the database access.   

\section{Description of the Keyphrase Cloud Generation}
The motivation to develop an automatic keyphrase extraction process is to create an hierarchical 3 layers representation of news. At the base level are the words provenient from ASR output, at the middle layer are the concepts (keyphrases), and at the top the index (topic). Our focus on this paper is on the middle or concept layer where every keyphrase represents a concept. Wherefore our concern is to identify the most relevant concepts without losing substantial precision. The cloud is generated because it provides a fast overview over the concepts of the most recent news. 

\subsection{Keyphrase Extraction}
\label{subsection:keyphraseExtraction}

Our Keyphrase extraction implementation was built over Maui-indexer toolkit~\cite{Medelyan2010}, state-of-art keyphrase extraction toolkit, which is respectively an extended version of the KEA~\cite{Witten1999}. 
The first step consisted of an adaptation of the toolkit for Portuguese, i.e., the incorporation of the Portuguese stemmer and the list of the stopwords based on the KEA porting work for Portuguese~\cite{Abadia2006}. 
The minimum number of occurrences of phrase in a news document to be consider as a keyphrase candidate is one. 

Malformed keyphrases are candidate phrases starting or ending or both in stop words. Hence, they are excluded.
Maui-indexer extracts several features: TF, IDF, TFxIDF, position of first occurrence, position of the last occurrence, distance between last and first occurrence, and number of words in the phrase. These are the baseline features (base) and the words are steamed. 
Then, we enriched the feature extraction process with the following features:
\begin{itemize}
\item number of characters (f1) -- empirically noun words that are long tend to be relevant; 
\item number of Named Entities (f2) -- recurrently named entities are important keyphrases; the number of named entities per keyphrase is obtained decomposing it into words and labelling them using the MorphoAdorner Name Recognizer\footnote{\url{http://morphadorner.northwestern.edu/}};
\item number of Capital letters (f3) -- the identification of acronyms is the main motivation to include this feature.
\item Part of Speech (POS) tags (f4) -- keyphrases are usually nouns or noun phrases, verbs or verb phrases are less frequent, and the remaining POS tags are rare;
\item probability of the keyphrase in a 4-ngram domain model (f5) -- feature included to capture how frequent it is to find in a larger BN corpus. 
\end{itemize} 
The 4-ngram domain model is interpolation of back-off n-gram models of BNs generated for AUDIMUS ASR language model~\cite{Martins2010}. It contains about 58,000 unigrams, 7,000,000 bigrams, 15,000,000 trigrams, and 10,000,000 4-grams. The model was compressed based on the Minimal Perfect Hash method developed for language models~\cite{Guthrie2010} to allow both faster access to the model and lower memory footprint. The smooth-nlp toolkit\footnote{\url{http://tinyurl.com/MphfCompres}} was used for this purpose. The compress model is about 12\% of the original size.

The training phrase consists of extracting features for every candidate keyphrases identified in the train set described in Section \ref{subsection:GoldStandardCorpus} and include a binary label stating whether it is a keyphrase. Then, the machine learning classifier, e.g. decision tree, establishes a association between the features values and the binary class. This  association, that can measured in terms of probability, is designated as classification model. The candidate phrases are ranked based on the probability generated by the model and it is selected top ranked phrases.

The machine learning classifier used by the Maui-indexer is the Bootstrap Aggregating (bagging)~\cite{breiman:1996} applied to C4.5 decision tree algorithm~\cite{quinlan:1993}.  Bagging is usually applied to decision tree models to reduce variance and mitigate overfitting by combining classification of randomly generated sample training sets. 

We decided to use the CART algorithm, that stands for Classification and Regression Tree, because it usually has better performance than the C4.5~\cite{Marujo:2010} and WEB\footnote{\url{http://tinyurl.com/C4-5VsCart}} . In addition, we wanted to improve the performance classification for small keyphrases extraction ($<= 10 $ keyphrases) -- to our knowledge the large majority of keyphrases extraction experiments fall into this category -- and discover for large keyphrases extraction ($>= 30 $ keyphrases) how well they perform.

\subsection{Cloud Generation}
\label{subsection:cloudGeneration}
The cloud is generated using the 10 top ranked keyphrases from about 10 top news from all tv/radio channels. Each cloud encapsulates the 20 most frequent keyphrases, i.e., each news is represented on average by 2 keyphrases. 
This procedure is used to generate both the all topics and the 12 topic based clouds. The only difference between the all topics cloud and the topic based clouds is the restriction of the top news that need belonging to the cloud topic.

At these stage of development, the top news are selected based on 2 heuristics: 
\begin{itemize}
\item time order -- news broadcast in beginning of the program are usually more relevant, i.e., they are usually order by their relevance. 
\item number of occurrences - the same relevant news are frequently transmitted in several channels simultaneously or within a temporal windows, e.g. 6 hours.
\end{itemize}

\subsection{Cloud Visualization}
\begin{figure}[h!]
\centerline{\mbox{\includegraphics[height=4cm]{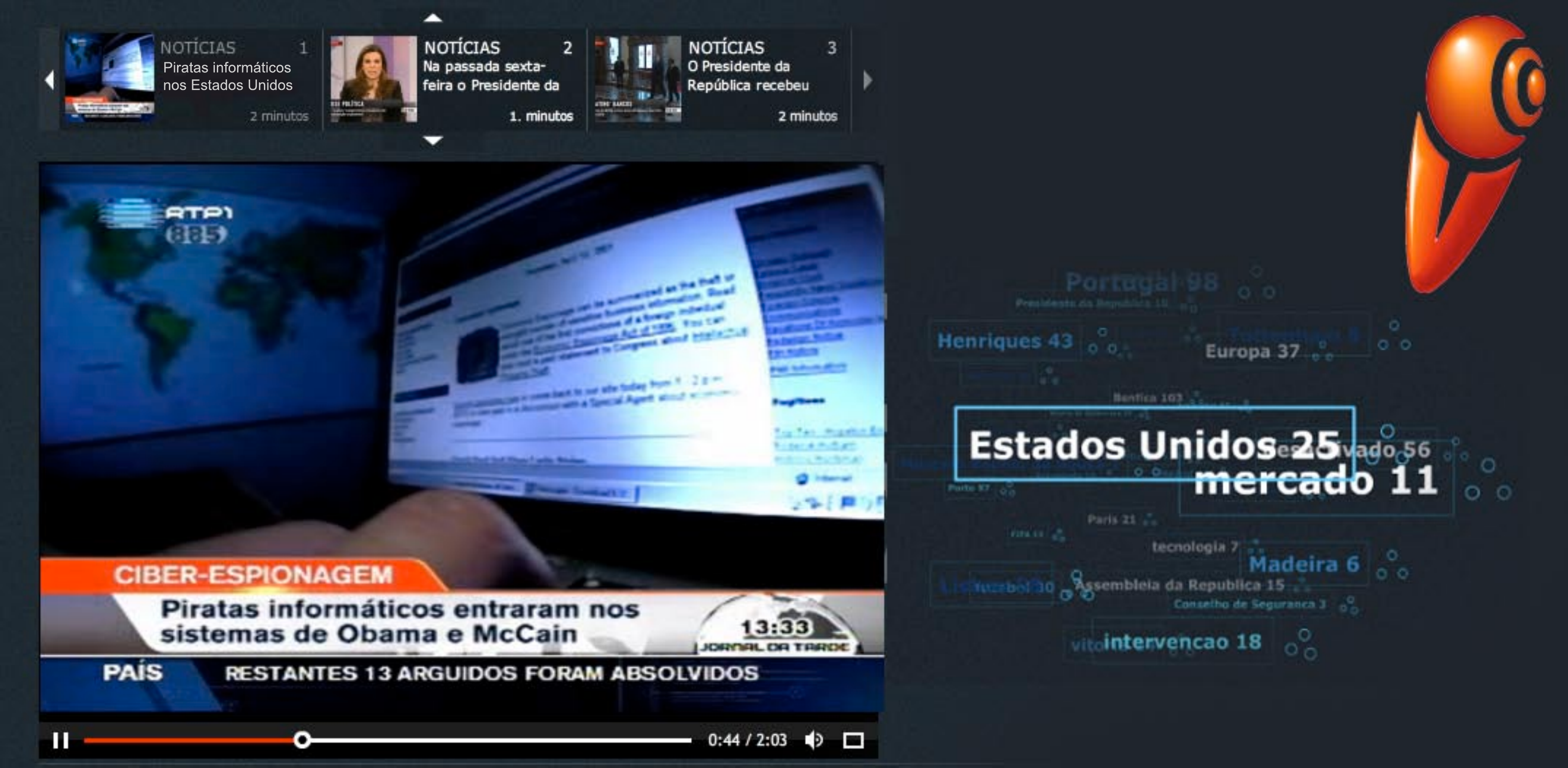}}}
\caption{MMS All Topic Cloud interface.}
\label{fig:MMS-Cloud_Interface}
\end{figure}

When a user logins into the system, he visualizes the generic 3D cloud (Figure \ref{fig:MMS-Cloud_Interface} on the right). The numbers in parenthesis reflect the total number of occurrences of the keyphrase in the news and it is an provisional way to show the relevance of the keyphrase in a 3D cloud.
\vspace{-0.3cm}

When the user clicks in one keyphrase of the cloud, the most recent news video containing the keyphrase is played. 
While the news video is being played, the user can also navigate horizontally through the playlist of news containing the keyphrase over time in all tv/radio channels. In addition, the user can also navigate vertically and visualize the different concepts.

\section{Results}

\subsection{Gold Standard Corpus}
\label{subsection:GoldStandardCorpus}

The gold standard corpus BN annotated with the respective keyphrases was developed because we did not find available any annotated corpus of the domain . Since we have to create a new corpus, we started to build for Portuguese, but in the future we will extend our approach to English.
Having this corpus was crucial to train and test our supervised extraction approach.

The gold standard is made of 8 BN programs, containing a total of 110 news subset, from the European Portuguese ALERT BN database~\cite{Neto2003}. The news transcription were produced by AUDIMUS, the in-house ASR for Portuguese, with low WER (14,56 \% on average); and punctuated and capitalize automatically using an in-house Capitalization and Punctuation module~\cite{Batista2009}. The programs were segmented automatically. Later, each news was manually reexamined to fix any segmentation error.  

Afterward, one annotator was asked to extract all keyphrases that summarize each news, that is, to mark all keyphrases that he assessed as needed to represent all concepts shown for each news. The maximum number of words in a keyphrase was limited to 5 words.  
This is another substantial difference from the large majority of keyphrase extraction work and stimulated the creation of gold standard. 
\begin{table}[h]
\centering
\begin{tabular}{|c|c|c|c|}
\hline
\bf{Type} &\bf{\#News}&\bf{Total \#Words}&\bf{Avg.\#Keyphrases}\\
\hline
\hline 
Train & 100 & 29225 & 24 \\
\hline
Test & 10 & 3896 & 29 \\
\hline
\end{tabular}
\caption{\label{corpusDescription} Corpus description.}
\end{table}
\vspace{-0.5cm}
\subsection{Keyphrase Extraction Evaluation}
To evaluate the performance of our keyphrase extraction method, the standard Information Retrieval metrics, i.e., precision (P), recall (R) and F1 measure were used.

The Tables \ref{KeyphraseExtractionResultsC4.5} and  \ref{KeyphraseExtractionResultsCART} shows the average results obtained in the test set. The leftmost column shows the maximum number of keyphrases extracted per news. The average number of keyphrases correctly identified appears in 3$^{rd}$ column in the middle.

The first rows of the Tables \ref{KeyphraseExtractionResultsC4.5} and  \ref{KeyphraseExtractionResultsCART} are our baseline results that are inline with reported state of art values for supervised approaches~\cite{Medelyan2006}. They were calculated for using the top 30 keyphrase candidates extracted. The following 10 rows are included to hightlight the performance gains per each new feature. They are followed by the combination of features. Only the best features combinations were included.     

\begin{table}[h]
\setlength{\tabcolsep}{1pt}
\centering
\begin{tabular}{|c|c|c|c|c|c|c|}
\hline
\bf{\# Keyphrases}&\bf{Features} & \bf{\#Keyphrases} & \bf{P} & \bf{R} &\bf{F1}\\
 \bf{Extracted} &  & \bf{Identified} &  & &\\
\hline
30 & base & 8.5 & 28.33 & 31.71 & 29.93 \\
\hline
\hline 
30 & base+f1 & 8.6 & 28.67 & 32.02 & 30.25\\
\hline
30 & base+f2 & 9 & 30 & 33.69 & 31.74 \\
\hline 
30 & base+f3 & 9.2 & 30.67 & 35.4 & 32.86 \\
\hline 
30 & base+f4 & 8.1 & 27 & 30.56 & 28.67\\
\hline 
30 & base+f5 & 8.9 & 29.67 & 33.73 & 31.57\\
\hline 
\hline 
30 & base+f1+f2& 8.6 & 28.67 & 32.02 & 30.25 \\
\hline
30 & all prev.+f3& 9.4 & 31.33 & 35.81 & 33.42\\
\hline
30 & all prev.+f4& 9.7 & 32.33 & 36.86 & \bf{34.45}\\
\hline
10 & all prev.+f4& 5 & 50 & 19.39 & 27.95\\
\hline 
\hline
10 & All & 5.3 & \bf{53} & 20.63 & 29.7 \\
\hline
20 & All & 7.4 & 37 & 28.21 & 32.01 \\
\hline
30 & All & 9.2 & 30.67 & 35.12 & 32.74 \\
\hline
35 & All & 10 & 28.57 & 37.79 & 32.54 \\
\hline
40 & All & 10.3 & 25.75 & 38.87 & 30.98\\
\hline
\end{tabular}
\caption{\label{KeyphraseExtractionResultsC4.5} Keyphrase Extraction results using Bagging + C4.5 decision tree}
\end{table}
\begin{table}[h]
\setlength{\tabcolsep}{1pt}
\centering
\begin{tabular}{|c|c|c|c|c|c|c|}
\hline
\bf{\# Keyphrases}&\bf{Features} & \bf{\#Keyphrases} & \bf{P} & \bf{R} &\bf{F1}\\
 \bf{Extracted} &  & \bf{Identified} &  & &\\
\hline
30 & base & 8.6 & 28.67 & 32.32 & 30.38 \\
\hline
\hline
30 & base+f1 & 9.4 & 31.33 & 35.48 & 33.28\\
\hline
30 & base+f2 & 9.3 & 31 & 35.34 & 33.03 \\
\hline
30 & base+f3 & 9 & 30 & 33.9 & 31.83 \\
\hline
30 & base+f4 & 9.1 & 30.33 & 34.55 & 32.3 \\
\hline
30 & base+f5 & 8.9 & 29.67 & 33.19 & 31.33 \\
\hline 
\hline 
30 & base+f1+f2 & 9.4 & 31.33 & 35.48 & 33.28 \\
\hline
30 & all prev.+f3 & 9.4 & 31.33 & 35.78 & 33.41 \\
\hline
30 &all prev.+f4& 9.4 & 31.33 & 35.39 & 33.24\\
\hline
10 & all prev.+f4 & 4.4 & 44 & 16.76 & 24.27\\
\hline 
\hline
10 & All & 4.6 & 46 & 17.33 & 25.18\\
\hline
20 & All & 7.1 & 35.5 & 26.86 & 30.58\\
\hline
30 & All & 9.4 & 31.33 & 36.19 & 33.59\\
\hline
35 & All & 9.8 & 28 & 37.68 & 32.13\\
\hline
40 & All & \bf{10.4}  & 26 & \bf{40.18} & 31.57\\
\hline
\end{tabular}
\caption{\label{KeyphraseExtractionResultsCART} Keyphrase Extraction results using Bagging + CART decision tree}
\end{table}
\vspace{-0.5cm}
\section{Conclusions}
This paper presents a novel method to extract (keyphrase extraction), represent (hierarchical 3 layer representation) and visualize (tag cloud) the semantic content of BNs transmitted in the last 6 hours, in a way that allows users to skim its content and to jump to the top relevant news.

Although our best keyphrase extraction results (higher F1 measure) are obtained using the C4.5 decision tree, in the large majority of the results, the CART outperforms the C4.5 using, for instance, the baseline features and extracting 30 keyphrases.

We also observed that the CART model is more robust to the inclusion of additional features because it gradually improves its performance, while the C4.5 classifier performance fluctuates. Another conclusion that can be draw from the results is that the C4.5 perform better in the extraction of few keyphrases and CART produces better results during the extraction of large number of keyphrases.

The news anchor names, that are said when a studio reporter talks with a field reporter was identified as an extra source of noise to the keyphrase extraction for large keyphrases extraction ($>= 30 $ keyphrases). Thus, future work should address this issue. Another enhancement to the keyphrase extraction that we hope to include in the future is to determine automatically how many keyphrase candidates should be retrieved per news document, using a linear interpolation of the keyphrase candidate confidence values with the news size.

\section{Acknowledgements}
The authors would like to thank David Matos, Hugo Meinedo, and Isabel Trancoso for their help. Support for this research by FCT through the Carnegie Mellon Portugal Program and under FCT grant SFRH/BD/33769/2009. This work was also partially funded by European Commission under the contract FP7-SME-262428 EuTV, QREN SI IDT 2525, and SI IDT 5108. Support by FCT (INESC-ID multiannual funding) through the PIDDAC Program funds.
\eightpt
\bibliographystyle{IEEEtran}
\bibliography{bibliography}

\begin{thebibliography}{10}
\providecommand{\url}[1]{#1}
\csname url@samestyle\endcsname
\providecommand{\newblock}{\relax}
\providecommand{\bibinfo}[2]{#2}
\providecommand{\BIBentrySTDinterwordspacing}{\spaceskip=0pt\relax}
\providecommand{\BIBentryALTinterwordstretchfactor}{4}
\providecommand{\BIBentryALTinterwordspacing}{\spaceskip=\fontdimen2\font plus
\BIBentryALTinterwordstretchfactor\fontdimen3\font minus
  \fontdimen4\font\relax}
\providecommand{\BIBforeignlanguage}[2]{{%
\expandafter\ifx\csname l@#1\endcsname\relax
\typeout{** WARNING: IEEEtran.bst: No hyphenation pattern has been}%
\typeout{** loaded for the language `#1'. Using the pattern for}%
\typeout{** the default language instead.}%
\else
\language=\csname l@#1\endcsname
\fi
#2}}
\providecommand{\BIBdecl}{\relax}
\BIBdecl

\bibitem{Meinedo2010}
H.~Meinedo, A.~Abad, T.~Pellegrini, J.~Neto, and I.~Trancoso, ``{The L2F
  Broadcast News Speech Recognition System},'' in \emph{Fala2010}, Vigo, Spain,
  2010.

\bibitem{Jones2003}
D.~Jones, F.~Wolf, E.~Gibson, E.~Williams, E.~Fedorenko, D.~Reynolds, and
  M.~Zissman, ``{Measuring the Readability of Automatic Speech-to-Text
  Transcripts},'' in \emph{Proceedings of Eurospeech}, 2003, pp. 1585--1588.

\bibitem{Amaral2008}
R.~Amaral and I.~Trancoso, ``{Topic Segmentation and Indexation in a Media
  Watch System},'' in \emph{Interspeech 2008}.\hskip 1em plus 0.5em minus
  0.4em\relax Brisbane, Australia: ISCA, 2008, pp. 2183--2186.

\bibitem{Kuo2007}
B.~Y.-L. Kuo, T.~Hentrich, B.~M.~. Good, and M.~D. Wilkinson, ``{Tag clouds for
  summarizing web search results},'' in \emph{Proceedings of the 16th
  international conference on World Wide Web - WWW '07}.\hskip 1em plus 0.5em
  minus 0.4em\relax New York, USA: ACM Press, 2007, pp. 1203--1204.

\bibitem{Tsagias2008}
M.~Tsagias, M.~Larson, and M.~de~Rijke, ``{Term clouds as surrogates for user
  generated speech},'' \emph{Proceedings of the 31st annual international ACM
  SIGIR'08}, p. 773, 2008.

\bibitem{Cohen1995}
J.~D. Cohen, ``{Highlights : Language- and Domain-Independent Indexing Terms
  for Abstracting Automatic},'' \emph{English}, vol.~46, no.~3, pp. 162--174,
  1995.

\bibitem{Luhn:1957}
H.~P. Luhn, ``A statistical approach to mechanized encoding and searching of
  literary information,'' \emph{IBM J. Res. Dev.}, vol.~1, pp. 309--317,
  October 1957.

\bibitem{Salton:1974}
G.~Salton, C.~S. Yang, and C.~T. Yu, ``A theory of term importance in automatic
  text analysis,'' Ithaca, NY, USA, Tech. Rep., 1974.

\bibitem{Chien:1997}
L.~Chien, ``Pat-tree-based keyword extraction for chinese information
  retrieval,'' in \emph{Proceedings of the 20th annual international ACM SIGIR
  conference on Research and development in information retrieval}, ser. SIGIR
  '97.\hskip 1em plus 0.5em minus 0.4em\relax New York, NY, USA: ACM, 1997, pp.
  50--58.

\bibitem{Ercan:2007}
G.~Ercan and I.~Cicekli, ``Using lexical chains for keyword extraction,''
  \emph{Information Processing \& Management}, vol.~43, no.~6, pp. 1705 --
  1714, 2007, text Summarization.

\bibitem{Zhang2006}
K.~Zhang, H.~Xu, J.~Tang, and J.~Li, ``{Keyword extraction using support vector
  machine},'' \emph{Advances in Web-Age Information Management}, pp. 85--96,
  2006.

\bibitem{Medelyan2006}
O.~Medelyan and I.~H. Witten, ``{Thesaurus based automatic keyphrase
  indexing},'' \emph{Proceedings of the 6th ACM/IEEE-CS joint conference on
  Digital libraries - JCDL '06}, p. 296, 2006.

\bibitem{Zhang2008}
C.~Zhang, H.~Wang, Y.~Liu, D.~Wu, Y.~Liao, and B.~Wang, ``{Automatic Keyword
  Extraction from Documents Using Conditional Random Fields},''
  \emph{Information Systems}, vol.~3, 2008.

\bibitem{Medelyan2010}
O.~Medelyan, V.~Perrone, and I.~H. Witten, ``{Subject metadata support powered
  by Maui},'' in \emph{Proceedings of the 10th annual joint conference on
  Digital libraries - JCDL '10}.\hskip 1em plus 0.5em minus 0.4em\relax New
  York, New York, USA: ACM Press, 2010, p. 407.

\bibitem{Neto2011}
J.~Neto, H.~Meinedo, and M.~Viveiros, ``A media monitoring solution,'' in
  \emph{ICASSP 2011 - Int. Conf. on Acoustics, Speech, and Signal Processing},
  Prague, Czech Republic, 2011.

\bibitem{Witten1999}
I.~Witten, G.~Paynter, E.~Frank, C.~Gutwin, and C.~Nevill-Manning, ``{KEA:
  Practical automatic keyphrase extraction},'' in \emph{Proceedings of the
  fourth ACM conference on Digital libraries}.\hskip 1em plus 0.5em minus
  0.4em\relax ACM, 1999, pp. 254--255.

\bibitem{Abadia2006}
M.~Abadia, L.~Dias, and M.~Malheiros, ``{Automatic Extraction of Keywords for
  the Portuguese Language},'' \emph{Computational Processing of the Portuguese
  Language}, pp. 204--207, 2006.

\bibitem{Martins2010}
C.~Martins, A.~Teixeira, and J.~Neto, ``{Dynamic language modeling for European
  Portuguese},'' \emph{Computer Speech \& Language}, vol.~24, no.~4, pp.
  750--773, Oct. 2010.

\bibitem{Guthrie2010}
D.~Guthrie, M.~Hepple, and W.~Liu, ``\BIBforeignlanguage{english}{Efficient
  minimal perfect hash language models},'' in
  \emph{\BIBforeignlanguage{english}{Proceedings of LREC'10}}.\hskip 1em plus
  0.5em minus 0.4em\relax Valletta, Malta: European Language Resources
  Association (ELRA), May 2010.

\bibitem{breiman:1996}
L.~Breiman, ``{Bagging predictors},'' \emph{Machine learning}, vol.~24, no.~2,
  pp. 123--140, 1996.

\bibitem{quinlan:1993}
J.~Quinlan, \emph{{C4. 5: programs for machine learning}}.\hskip 1em plus 0.5em
  minus 0.4em\relax Morgan Kaufmann, 1993.

\bibitem{Marujo:2010}
L.~Marujo, ``{Voting Combination of Sentences Splitting Classifiers Applied to
  Several Types of Texts},'' \emph{Technical report}, 2010.

\bibitem{Neto2003}
J.~Neto, H.~Meinedo, R.~Amaral, and I.~Trancoso, ``{A system for selective
  dissemination of multimedia information resulting from the ALERT project},''
  in \emph{Proc. ISCA ITRW on Multilingual Spoken Document Retrieval}, Hong
  Kong, China, March 2003.

\bibitem{Batista2009}
F.~Batista, I.~Trancoso, and N.~Mamede, ``{Automatic Recovery of Punctuation
  Marks and Capitalization Information for Iberian Languages},'' in \emph{I
  Joint SIG-IL/Microsoft Workshop on Speech An Language Technologies for
  Iberian Languages}, Porto Salvo, Portugal, 2009, pp. 99--102.

\end{thebibliography}
\end{document}